\documentclass[reprint,aps,floats,floatfix,amsmath,amssymb,amsfonts,nofootinbib,longbibliography,superscriptaddresst]{revtex4-1}
\usepackage{graphicx}
\usepackage{dcolumn}
\usepackage{bm}
\usepackage{tensor} 
\usepackage{natbib}
\usepackage{subcaption}
\usepackage{float}

\usepackage[hyperindex,colorlinks]{hyperref}

\begin{document}

\title{Constraining the anomalous coupling of gravitational waves with double pulsar }

\author{Diego Santos de Jesus}
\email{diego.s.jesus@edu.ufes.br}
\affiliation{%
N\'ucleo Cosmo-ufes \& Departamento de F\'isica,  Universidade Federal do Esp\'irito Santo (UFES)\\
Av. Fernando Ferrari, 540, CEP 29.075-910, Vit\'oria, ES, Brazil.}%

\author{Hermano Velten}%
\email{hermano.velten@ufop.edu.br}
\affiliation{%
Departamento de F\'isica, Universidade Federal de Ouro Preto (UFOP), Campus Universit\'ario Morro do Cruzeiro, 35.400-000, Ouro Preto, Brazil}%

\author{Federico Piazza}%
\email{piazza@cpt.univ-mrs.fr}
\affiliation{%
Aix Marseille Univ, Universit\'{e} de Toulon, CNRS, CPT, Marseille, France}%

\date{\today}

\begin{abstract}
We revisit the decay of the orbital period in binary systems that occurs due to the emission of gravitational waves in the context of modified gravity models where the coupling $G_{gw}$ between matter and on-shell gravitons is allowed to differ from the Newton constant $G_N$. Using the most precise orbital parameters of binary pulsars, those of the Double Pulsar, we constrain the ratio $G_{gw}/G_N$ to the level of $10^{-4}$, improving by two orders of magnitude the present bound on this quantity.
\end{abstract}

\maketitle

\section{Introduction}

One desirable feature of alternatives to general relativity (GR) is being able to account for the acceleration of the universe. At smaller scales, however, these models should conform to the strict solar system tests and the wealth of observations that are coming from our own galaxy.  Popular models of modified gravity include a dynamical scalar field mediator on top of the metric field of GR.  

Scalar tensor theories of the Galileon type~\cite{Horndeski:1974wa,Nicolis:2008in} and their subsequent generalizations~\cite{Gleyzes:2014dya,Langlois:2015cwa} come equipped with (\emph{Vainshtein}) screening mechanisms~\cite{Babichev:2013usa} that suppress the contributions of the scalar mediator inside overdense regions. Screening, however, cannot make these models perfectly identical to GR.

For a large class of theories, the timelike gradient of the scalar field that evolves in time on cosmological scales persists \emph{locally}, inside virialized environments, thus \emph{piercing the Vainshtein screen}~\cite{jpv}.
One consequence of this is that the anomalous gravitational waves speed $c_T$ that these models can exhibit survives the local overdense environment of our galaxy. So the effect can be directly constrained by comparing the arrival times of the electromagnetic and gravitational signals emitted by a common source. This simple comparison has produced the impressive bound~\cite{monitor}
\begin{equation}
    -3\times 10^{-15}\leq \frac{c_T - c}{c}\leq 7\times 10^{-16}\,
\label{cTLigo}\end{equation}
and killed a large class of theories~\cite{Creminelli:2017sry}. Another victim of such a persistent timelike gradient is the very mechanism of \emph{self-acceleration} (see also~\cite{Lombriser:2016yzn} on this). The cosmological constant and all ``quintessence-like" models of dark energy drive the acceleration by contributing to the total energy momentum tensor with a relevant negative pressure component. Models of modified gravity offer a conceptual alternative to that, as the conformal factor relating the (physical-) Jordan and Einstein frames (see e.g.~\cite{Nitti:2012ev}) can be varying in time in such a way to contribute to accelerate the cosmological expansion rate in the physical frame. Such a variation, however, is effectively indistinguishable~\cite{Babichev:2011iz} from that of the Newton constant. The latter is constrained to the level of $\dot G_N/G_N \lesssim 10^{-2}H_0$ by Lunar Laser Ranging experiments~\cite{Williams:2004qba}, two orders of magnitude smaller than needed to have efficient self-acceleration.

Another potential signature of \emph{degenerate higher order theories}~\cite{Gleyzes:2014dya,Langlois:2015cwa}---even those that survive the above constraints~\cite{Crisostomi:2019yfo}---is that the coupling $G_{gw}$ between the Lagrangian responsible for describing gravitational waves with the matter fields is different than $G_N$ as defined, for example, by the binding gravitational energy of two objects. Currently, thanks to decades of monitoring, pulsars are the most precise astronomical tool to probe any eventual discord between $G_{gw}$ and $G_{N}$. By using the observed decay of the orbital period of the Hulse-Taylor pulsar~\cite{Weisberg:2010zz}, and assuming that $c_T=c$, the authors of~\cite{jpv} gave a constrain on the ratio $G_{gw}/G_N$ of the order $10^{-2}$. In this note we improve on that constraint by about other two orders of magnitude thanks to the newly released data from the Double Pulsar~\cite{kramer2}. 

\section{The double pulsar data}

Pulsars are excellent laboratories for testing relativistic predictions. Gravitation tests with pulsars are possible using a technique called pulsar timing. With such approach one has verified that the post-Keplerian parameters measured from a certain binary system match the predictions of the corresponding theory of gravity, but also place bounds on the magnitudes of parameters that arise in modified gravity theories \cite{stairs}.

The first pulsar has been observed in 1967 \cite{hewish}, but the first binary system containing such object, the PSR 1913+1916, named as the Hulse-Taylor pulsar, became a smooking gun for the confirmation of gravitational waves according to the GR predictions \cite{hulse-taylor}. The latter is related to the decay of the orbital period, presumed to occur as the system was losing energy in the form of gravitational waves. 

The periodic emission of their pulses makes pulsars excellent clocks and, via the pulsar timing technique, the orbital parameters can be obtained with great accuracy.

For binary systems, the pulsar mass $m_p$ and its companion $m_c$ are the two free parameters to be determined. In practise, the task is to determine the dependence of the Keplerian orbital parameters and the relativistic dependent post-Keplerian  parameters on the masses $m_p$ and $m_c$. The so called Damour-Deruelle post-Keplerian parameters that are more relevant here, because more constraining, are the advance of periastron
\begin{equation}
    \dot{\omega} = 3\frac{G_{N}^{2/3}}{c^{2}}\left ( \frac{P_{b}}{2\pi} \right )^{-5/3}\left (1-\epsilon^{2}  \right )^{-1}\left ( m_{p}+m_{c} \right )^{2/3}\ ,
\end{equation}
the amplitude of the Einstein delay
\begin{equation}
    \gamma  =  \epsilon\left ( \frac{P_{b}}{2\pi} \right )^{1/3}\frac{G_{N}^{2/3}}{c^{2}}\frac{m_{c}\left ( m_{p}+2m_{c} \right )}{\left ( m_{p}+m_{c} \right )^{4/3}}\ ,
\end{equation} 
and the Shapiro delay shape
\begin{equation}
     s = \frac{G_{N}^{-1/3}}{c}\left ( \frac{P_{b}}{2\pi} \right )^{-2/3}\,\frac{\left (m_{p} + m_{c}  \right )^{2/3}}{m_{c}}x  \, ,
\end{equation}
where the eccentricity $\epsilon$ and the orbital period $P_b$ are the Keplerian parameters and $x$ is the semi-major axis projection of the orbit. We have mentioned only the parameters that will be relevant for our analysis in this work. For a complete list see \cite{DamourDe}.

Whereas the above mentioned post-Keplerian parameters are related to the gravitational sector, the orbital decay rate of the binary system $\dot{P}_b$ is directly related to the radiative sector i.e., the emission of gravitation waves. In GR it reads
\begin{equation}\label{decaimentoPBMODI}
    \begin{aligned}
        \dot{P_{b}} |_{GR} = -  \frac{192\pi}{5}\frac{G_{N}^{5/3}}{c^{5}}\left ( \frac{P_{b}}{2\pi} \right )^{-5/3}\left[\frac{m_{p}m_{c}}{\left ( m_{p}+m_{c} \right )^{1/3}}\right] f(\epsilon)\ ,
\end{aligned}
\end{equation}
where the eccentricity dependent function $f(\epsilon)$ is 
\begin{equation}
 f(\epsilon)=   \left ( 1-\epsilon^{2} \right )^{7/2}\left ( 1 + \frac{73}{24}\epsilon^{2} + \frac{37}{96}\epsilon^{4} \right )\ .
\end{equation}

For a given orbit with well determined Period $P_b$ and eccentricity $\epsilon$, the post-Keplerian parameters will depend on the unknown masses $m_p$ and $m_c$. For example, measurements of $\dot{\omega}$ are able to constrain the total mass of the binary system $m_p+m_c$. Since the inferred measurement for each individual test can be explained by a certain combination of the masses $m_p$ and $m_c$, all tests will be satisfied simultaneously only with a unique set of masses values. Plotting the post-Keplerian tests in a $m_p \times m_c$ diagram, the crossing of all individual post-Keplerian tests will provide a common region in the $m_p, m_c$ parameter space where all in test are satisfied simultaneously yielding to the $m_p$ and $m_c$ stellar mass values.  For example, with accurate measurements of $\dot{\omega}$ and $ \gamma$, the masses in the binary the Hulse-Taylor system have been assessed in Refs. \cite{Weisberg:2010zz, weisberg} providing for this system $m_p=1.438(1) M_{\odot}$ and $m_c=1.390(1) M_{\odot}$.    

Though there are many binary pulsars systems, one of the most relevant is the Double Pulsar system, or PSR J0737 3039 AB. Both objects are visible as pulsars, it is relatively close and its orbital inclination relative to us is about half a degree from 90 degrees, leading to optimal conditions to measure relativistic effect as the pulsar's signals pass through the orbital plane, where they are most strongly influenced by the curvature of spacetime generated by the system itself \cite{shao}.
The Double pulsar has improved the accuracy of most previous gravity tests by orders of magnitude \cite{kramer1} and also allowing for new tests \cite{kramer2}.


In Table \ref{Table1} the orbital parameters for both systems are compared. All uncertainties are quoted at $1 \sigma$ level. It is clear that the Double pulsar has better accuracy in all parameters. Remarkably, the Shapiro delay parameter $s$, which was absent in the Hulse-Taylor data in 2010 \cite{Weisberg:2010zz}, but has been badly constrained in the 2016 analysis \cite{weisberg}, is the parameter responsible for setting the masses in the Double pulsar system.


\begin{table}[htb]
\centering
\caption{Orbital parameters to Double Pulsar and Hulse-Taylor pulsar taken from \cite{kramer2} and \cite{weisberg}.}
\label{Table1}
\begin{tabular}{cccc}
\hline \hline
Parameter                     & Double Pulsar \cite{kramer2}                    & Hulse-Taylor pulsar \cite{weisberg}\\ \hline 
$P_{b}$ (days)                & 0.1022515592973(10)              & 0.322997448918(3)  \\ [2pt]
$\epsilon$                    & 0.087777023(61)                  & 0.6171340(4)\\ [2pt]
$\dot{\omega}$ (deg/yr)       & 16.899323(13)                    & 4.226585(4)\\ [2pt]
$\gamma$ (ms)                 & 0.384045(94)                     & 4.307(4)\\ [2pt]
$\dot{P}_{b}$                 & -1.247920(78)$\times 10^{-12}$   &  -2.423(1) $\times 10^{-12}$ \\ [2pt]
$s$                           & 0.999936(+9/-10)                 & 0.68(+10/-6)         \\[2pt] 
$\dot{P}_{b}^{\text{ext}}$    & -1.68(+11/-10)$\times 10^{-16}$  &  -0.025 (4)$\times 10^{-12}$\\ [2pt] \hline
\end{tabular}
\end{table}

In order to correctly compute the decay of the orbital period one has to add external contributions $\dot{P}_b^{ext}$ since the pulsar frame is accelerated with respect to the solar system barycenter. $\dot{P}_{b}$ is affected not only by the emission of gravitational radiation, but also by two relevant effects namely, the radial acceleration due to the transverse motion of the binary pulsar, called the Shklovskii effect, and the physical radial acceleration due to the galactic gravitational potential. Thus, $\dot{P}^{\text{ext}}_{b} = \dot{P}^{\text{Shk}}_{b} +\dot{P}^{\text{gal}}_{b}$ and therefore the period decay observed is given by the difference between $\dot{P}_{b}$ and $\dot{P}^{\text{ext}}_{b}$.

For the application we have in mind, following the same strategy as presented in Ref. \cite{jpv}, the magnitude of the uncertainty in the $\dot{P}_b$ measurement is the key quantity to constrain $G_{gw}$. Now, with the Double pulsar, this uncertainty is about two orders of magnitude smaller than the Hulse-Taylor pulsar. Indeed, the distance to the pulsar is an important parameter in many applications of pulsar timing, especially when testing the effects of emitting gravitational waves. For the Hulse-Taylor pulsar, the limited accuracy of the distance precludes any improvement as a gravity experiment, in \cite{Deller-Weisberg:2018}, the parallax measurement for the Hulse-Taylor pulsar is relatively low ($\sim 3\sigma$) which prevents an improved test of general relativity using orbital decay. As the Double Pulsar is relatively close, there is a small but measurable curvature in front of the signal, which allows to measure its distance with greater precision and, consequently, also precisely the decay of the orbital period \cite{kramer2}.


\begin{figure}[htb]
    \centering
    \includegraphics[scale=.2]{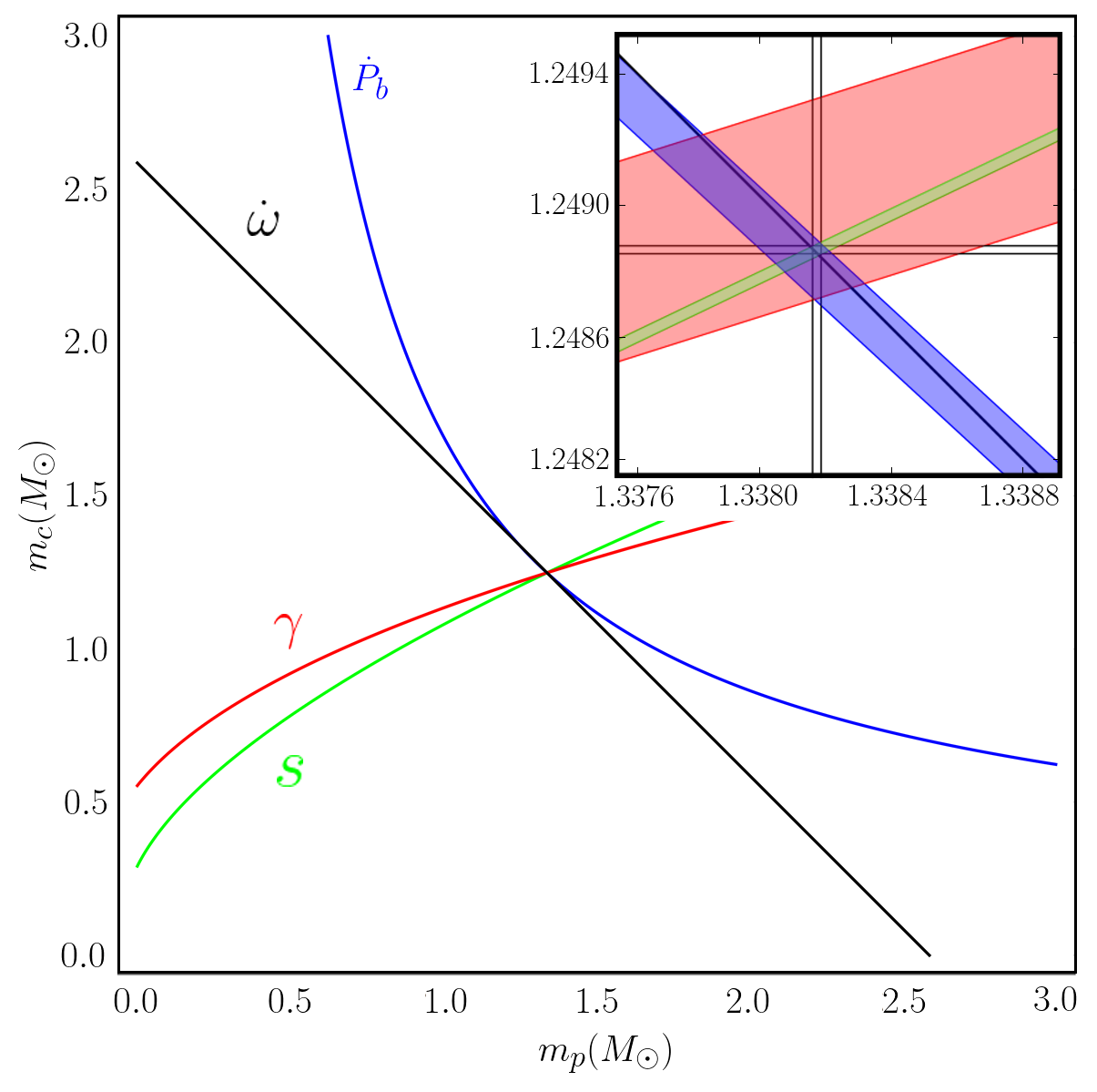}
    \caption{Mass-mass diagram for Double Pulsar.} 
     \label{Fig1}
\end{figure}


We show in Fig. \ref{Fig1} a reproduction of the mass-mass diagram for the Double pulsar as in Ref. \cite{kramer2}. From the crossing of the $\dot{\omega}$ and $s$ curves one obtains for the Double pulsar $m_{p} = 1.338185$(+12/-14) $M_{\odot}$ and $m_{c} = 1.248868$(+13/-11) $M_{\odot}$. The inset in this figure shows an amplification of the crossing region where the masses are indicated with the vertical and horizontal black lines. Such values are within the allowed region for the measurements of $\gamma$ and $\dot{P}_b$.

\section{Double pulsar bounds on $G_{gw}$}

In the presence of a mismatch $G_{gw} \neq G_N$, and under the hypothesis that the Keplerian orbits of the system are entirely governed by $G_N$, one can trace the required modification to the standard formula of the period decay rate for modified gravity theories allowing a different coupling between matter fields and the gravitational radiative sector. Then, by calculating the quadrupole formula for generic modified gravity scenarios in which such coupling constants of the gravitational sector $G_{gw}$ and the potential sector $G_{N}$ may differ from each other. Ref \cite{jpv} obtained for the period decay formula 
\begin{equation}\label{decaimentoPBMODI}
    \begin{aligned}
        \dot{P_{b}}|_{MG} = - \left(\frac{G_{gw}}{G_{N}}\right) \, \dot{P_{b}} |_{GR} .
\end{aligned}
\end{equation}
As presented in Ref. \cite{jpv}, the above formula can also have an extra factor $c/c_T$ in the right hand side but in light of the constraints obtained in \ref{cTLigo} one has $c/c_T=1$. 

 With the orbital parameters of the double pulsar we can get the best range for the $G_{gw} / G_{N}$ ratio, as compared to the values obtained by \cite{jpv}, the decay value of the orbital period due to the emission of gravitational waves is about 25 times more accurate than in the Hulse-Taylor pulsar \cite{weisberg}, which is currently limited by uncertainties in correcting for external effects.

\begin{figure}[H]
    \centering
    \includegraphics[scale=.2]{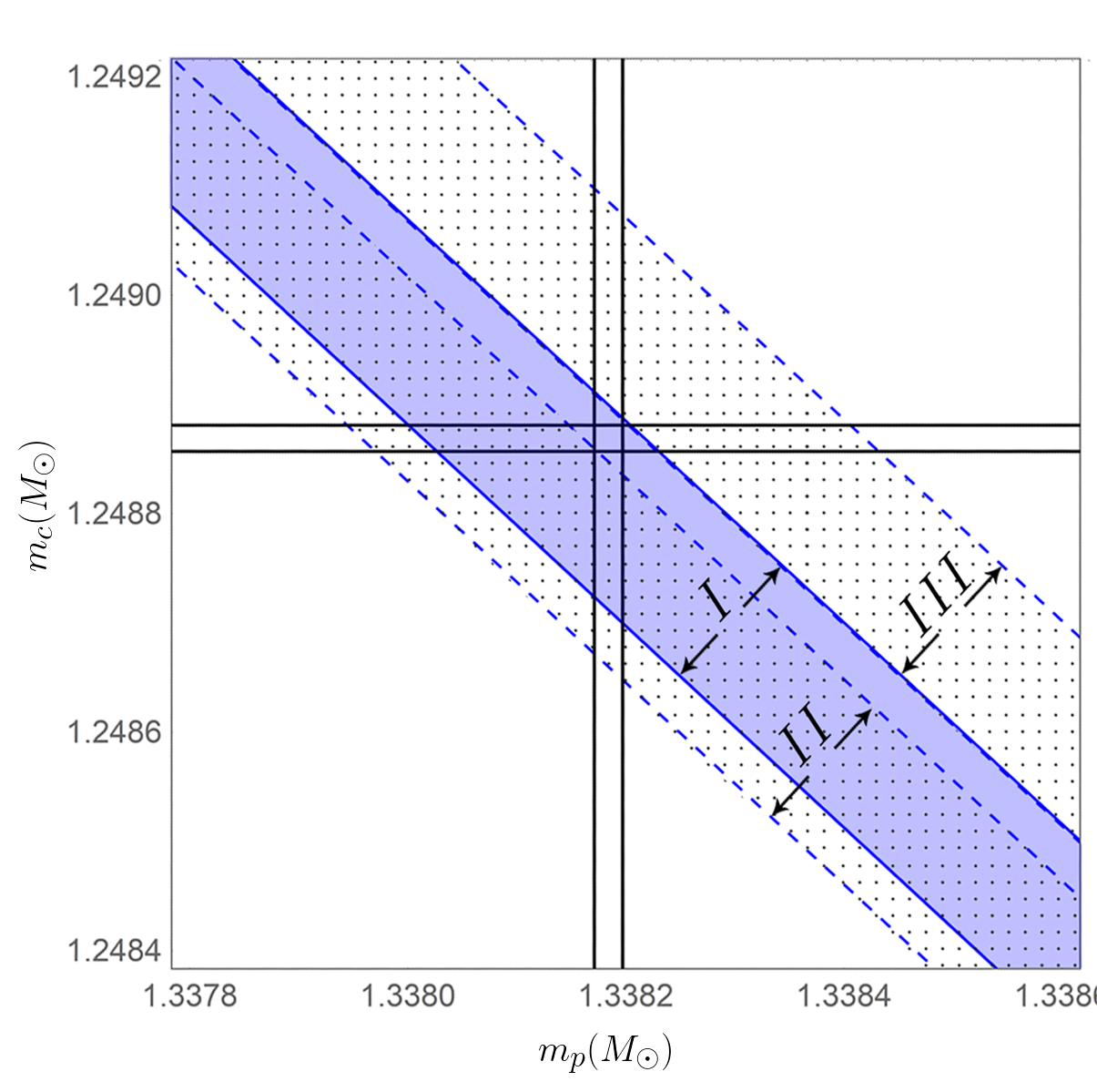}  
       \caption{Allowed variation of $\dot{P}_{b}$ (blue dashed) to determine $G_{gw}/G_{N}$. The region $I$ in blue is the values of $\dot{P}_{b}$, where we have that $G_{gw} = G_{N}$, the dotted areas are the maximum and minimum limits for the relation studied, where close to the blue area, we have that the maximum limit allowed (region $II$) is such that $G_{gw} = 1.00007 G_{N}$ and for the minimum limit (region $III$), further out of the blue area, this limit is $G_{ gw} = 0.99975 G_{N}$.}\label{DMMRESULT}
\end{figure}

Let us now proceed with the same strategy as in Ref. \cite{jpv}. To determine $G_{\text{gw}} / G_{N}$ we vary \eqref{decaimentoPBMODI} to the maximum limit that is determined by the area where the masses meet.  In Fig. \ref{DMMRESULT} the blue shaded corresponds to the standard GR result i.e., $G_{gw}/G_{N}=1$. By letting $G_{gw}/G_{N}\neq1$ values this shaded region move diagonally to the upper right  (bottom left) direction if $G_{gw}/G_{N} <1 (>1)$. Our criteria is such that the shaded region can not move beyond the crossing of the horizontal and vertical black lines determining the masses of the system. In the figure the maximum allowed limits will correspond to regions $II$ and $III$. Thus, this variation is such that the allowed range of values for $G_{ \text{gw}} / G_{N}$ becomes
\begin{equation}
   25 \cdot 10^{-5} \ \leq \ \frac{G_{gw}}{G_{N}} - 1 \ \leq \ 7 \cdot 10^{-5} \ ,
\end{equation}
as obtained from the diagram in Figure \ref{DMMRESULT}. Of course, such limits are in agreement with the Einstein delay curve.

\section{Conclusions}

We have improved by two orders of magnitudes the existing bounds on $G_{gw}/G_{N}$. This has been made possible by the newly released  data of the double pulsar, in which the uncertainty on the orbital period decay has reduced to 0.013\%. This constitutes an impressive improvement with respect to that of the Hulse-Taylor pulsar, which was measured with uncertainties in $\dot{P}_b \sim 0.3\%$ .

The allowed parameter space of modified gravity theories is rapidly shrinking and part of the original motivations for these models (i.e. a mechanism of self-acceleration qualitatively different than that of a standard negative-pressure component) seem now incompatible with data, as mentioned in the Introduction. 

A nice updated review of the current constraints on DHOST theories~\cite{Langlois:2015cwa} is given in~\cite{Crisostomi:2019yfo}, directly in terms of the parameters of the effective field theory formalism for cosmological perturbations~\cite{Piazza:2013coa}. Of the parameters governing cubic and higher order operators, after the existing constraints are applied,  only one parameter survives, which the authors of~\cite{Crisostomi:2019yfo} denote $\beta_1$. Such a parameter is constrained directly by the ratio $G_{gw}/G_N$. The present work improves the bound on $\beta_1$ from $\beta_1\lesssim 10^{-2}$ to $\beta_1\lesssim 10^{-4}$. 

Despite the difficulties of scalar tensor theories, the standard $\Lambda$CDM model of cosmology is not without issues on its own, with the more or less severe cosmological tensions  that have been emerging in the last decade or so  (see e.g.~\cite{Perivolaropoulos:2021jda}). These problems, however,  seem somewhat more subtle and puzzling than what modified gravity seems able to address at present. 

Finally, it is curious to note that the bound that we have obtained on $G_{gw}/G_N$ is not far from the precision with which $G_N$ itself is measured. Despite the experimental efforts since Cavendish's time, the accuracy on the measured value of $G_N$ is still poor compared to other fundamental constants. As reported by Ref. \cite{NatureLi}, up to date measurements of $G_N$ using torsion pendulum experiments have relative standard uncertainties $\sim 11$ parts per million i.e., $\mathcal{O}(10^{-5})$.  
\begin{acknowledgments}

The authors thank Filippo Vernizzi for discussions and FAPEMIG/FAPES/CNPq/CAPES and Proppi/UFOP for financial support. 
\end{acknowledgments}

\end{document}